\documentclass[11pt,twoside]{article}

\usepackage{asp2004}
\usepackage{epsf}
\usepackage{psfig}
\usepackage{lscape}

\begin{document}
\title{Solution of NLTE Radiative Transfer Problems Using
Forth-and-Back Implicit Lambda Iteration}   
\author{Olga Atanackovi\'c-Vukmanovi\'c}   
\affil{Department of Astronomy, Faculty of Mathematics, University
of Belgrade, Studentski trg 16, 11000 Belgrade, Serbia and
Montenegro}

\begin{abstract} The basic idea of an extremely fast convergent
iterative method, the Forth-and-Back Implicit Lambda Iteration
(FBILI), is briefly described and the applications of the method
to various RT problems are listed and discussed.
\end{abstract}


\section{Introduction}

Radiative transfer (RT) is the underlying physical phenomenon in
many astrophysical problems and among the most difficult to deal
with. The main difficulty arises from the non-local and in general
non-linear coupling of the radiation field and the state of the
gas.

In the problem of NLTE line formation in a given atmospheric
model, the internal state of the gas depends, via radiative
transitions, on the radiation field intensity, which in turn
depends, via the RT process, on the state of the gas over a wide
range of distant points. Mathematically, the non-local coupling is
performed by the simultaneous solution of the corresponding RT
equation describing the dependence of the mean intensity on the
source function $J=\Lambda [S]$ by means of the so-called $\Lambda
$ operator and the statistical equilibrium (SE) equations defining
the source function in terms of the mean intensity of the
radiation field $S=S(J)$.

In the well-known two-level-atom line formation problem the
non-local coupling is linear and the problem can be easily solved
by using either direct or iterative methods. In a more general
multilevel case, RT problem is non-linear and, therefore, an
iterative method is required.

The most straightforward iterative procedure, the so--called
$\Lambda $ iteration, solves radiative transfer and statistical
equilibrium equations in turn. However, in most cases of interest
(in scattering dominated media of large optical thickness) the
rate of convergence of this simple procedure is infinitely slow.

A broad class of ALI (Approximate (or Accelerated) Lambda
Iteration) methods, currently in use, is based on the use of
certain physical or computational approximations of the $\Lambda $
operator within an iterative procedure. These methods usually
employ some free parameter controlling the convergence and almost
always need additional acceleration by some mathematical
techniques (Ng acceleration, successive over-relaxation method,
etc.) to achieve high convergence rate (\citeauthor{hub03}
\citeyear{hub03};\ \citeauthor{av04} \citeyear{av04}).

\citet*{acs97} developed the Forth-and-Back Implicit Lambda
Iteration (FBILI) - a simple, accurate and extremely fast
convergent method to solve NLTE RT problems. FBILI dramatically
accelerates the convergence of the classical $\Lambda $ iteration
while retaining its straightforwardness. In this paper the basic
idea of the method is briefly explained and the applications to
various RT problems are shown and discussed.

\section{The Idea of FBILI Method}

In order to demonstrate the basis of the FBILI method we shall
consider the well-known case of the two-level atom line formation
(with complete redistribution and no overlapping continuum) in a
plane-parallel and static atmosphere. Under these assumptions, the
specific intensity of the radiation field $I_{x\mu }(\tau )$ is
described by the RT equation of the form:
\begin{equation}
\mu {{dI_{x \mu} }\over {d\tau }} = \varphi _x [I_{x \mu }(\tau )
-S(\tau )] \ \ ,
\end{equation}
where $\tau $ is the mean optical depth, $x$ is the frequency
displacement from the line center in Doppler width units, $\mu $
is the cosine of the angle between the photon's direction and the
outward normal and $\varphi_x$ is the absorption-line profile,
normalized to unity. The line source function (SE equation for a
two-level atom) has the following form:
\begin{equation}
S(\tau )=\varepsilon B(\tau ) + (1-\varepsilon )J_\varphi (\tau )\
\ ,
\end{equation} where $\varepsilon $ is the standard NLTE
parameter representing the branching ratio between the thermal
(LTE) contribution $B(\tau )$ and the scattering term
\begin{equation}
J_\varphi (\tau )= \int J_x \varphi _x dx = {1\over 2}\int _{-1}^1
d\mu \int _{-\infty }^{\infty } dx \varphi _xI_{x \mu }(\tau )\ \
,
\end{equation}
which accounts for the angle and frequency coupling of the
specific intensities at the given depth point $\tau $.

The basic idea of the forth-and-back implicit $\Lambda $ iteration
in the solution of the problem is as follows. First, as suggested
by the existence of two separate boundary conditions, the FBILI
uses a separate description of the propagation of the in-going
intensities of the radiation field $I^-_{x \mu }(\tau )$ with
initial conditions at the surface ($\tau =0$) and of the out-going
intensities $I^+_{x \mu }(\tau)$ with initial conditions at the
bottom of the atmosphere ($\tau =\tau _N$). Second, although the
values of the radiation field are unknown, its propagation can be
easily represented by using the integral form of the RT equation
and assuming polynomial (e.g. piecewise quadratic) representation
of the source function between two successive depth points. Thus,
for each depth point one can write linear relations for the
specific intensities as functions of the unknown values of the
source function and of its derivative. Following the idea of
iteration factors, it is the iterative computation of the
coefficients of these {\it implicit} relations (implicit, as the
source function is a priori unknown), rather than that of the
unknown functions themselves, which greatly accelerates the
convergence of the direct iterative scheme.

In the first part of each iteration (forward process), proceeding
from the upper boundary condition, using the integral form of the
RT equation for the in-going intensities for each layer ($\tau
_{l-1}, \tau _l$):
\begin{equation}
I^-_{x\mu }(\tau _l) = I^-_{x\mu }(\tau _{l-1})e^{-\Delta \tau _l
\varphi _x/\mu } + \int _{\tau _{l-1}}^{\tau _l} S(t) e^{-(\tau
_l-t)\varphi _x/\mu} {{\varphi _x}\over {\mu }}dt \ \ ,
\end{equation}
where $\Delta \tau _l=\tau _l-\tau _{l-1}$, and assuming parabolic
behavior for the source function $S^\prime (\tau _{l-1})=
2{{(S(\tau _l)-S(\tau _{l-1}))}/{\Delta \tau _l}}- S^\prime (\tau
_l)$, one can write the linear {\it local implicit} relation
\begin{equation}
I^-_{x\mu }(\tau _l) = [{{a^-_{x\mu }}\over {S^o(\tau _{l})}} +
b^-_{x\mu }] S(\tau _l) + c^-_{x\mu } S^{\prime} (\tau _l)\ ,
\end{equation}
representing the values of the in-going intensities $I_{x
\mu}^-(\tau _l)$ at a given optical depth point $\tau _l$ in terms
of yet unknown values of the source function $S(\tau _l)$ and of
its derivative $S^{\prime} (\tau _l)$. Here, $a^-_{x\mu }=
I^-_{x\mu }(\tau _{l-1})e^{-\Delta \tau _l\varphi _x/\mu
}+qS^o(\tau _{l-1})$ is computed with the old (known from the
previous iteration) source function $S^o$, whereas $b^-_{x\mu }$
and $c^-_{x\mu }$ depend only on the known optical distance
$\Delta \tau _l$. By integrating (5) over all frequencies and
directions we obtain the linear relation
\begin{equation}
J^-_\varphi (\tau _l) = b^-_l S(\tau _l) + c^-_l S^{\prime} (\tau
_l)\
\end{equation}
representing {\it implicitly} the value of the in-going mean
intensity. Thus, in the forward process, we differ from the
classical $\Lambda $ iteration that re-calculates $J^-_\varphi $
from the old (known) source function $S^o(\tau )$, in using the
old source function to compute, at each optical depth point $\tau
_l (l=1,N)$, the coefficients $b^-_l$ and $c^-_l$ of the linear
relation (6). The coefficients are stored for further use in the
backward process of computation of the new values of $S(\tau )$.
Let us note here that the ratio in Eq. (5) of the non-local part
of the in-going intensity $a^-_{x\mu }$ to the current source
function $S^o(\tau _l)$ is actually the iteration factor. Since it
is the only information that is carried from the previous
iteration step, an extremely fast convergence is to be expected.

The final aim is to derive, at each optical depth point $\tau _l$,
an {\it implicit} linear relation between the 'full' mean
intensity and the source function
\begin{equation}
J_\varphi (\tau _l) =a_l + b_l S(\tau _l)\ \end{equation} that,
together with equation (2), leads to the new source function. To
obtain this, we need the coefficients of the corresponding
relation for the out-going mean intensity.

In the backward process we proceed from the bottom where $I^+_{x
\mu }(\tau _{N})$, i.e. $J_\varphi ^+(\tau _N)$ is known or more
precisely, the coefficients of the {\it implicit} relation for the
out-going specific intensities and, therefore, for the out-going
mean intensity
\begin{equation}
J^+_\varphi (\tau _l) = a^+_l + b^+_lS(\tau _l)\ \ ,
\end{equation} are known at $\tau =\tau _N$. Eliminating
$S^{\prime }(\tau _N)$ from Eq. (6) \citep[see][]{acs97} we obtain
the linear implicit relation (7), which together with SE equation
(2) leads to the new value of $S(\tau _N)$. The new values of
$S(\tau _{N})$, $S^\prime (\tau _{N})$ and, hence, $I^+_{x \mu
}(\tau _{N})$ are then used to compute the coefficients of the
linear relation (8) in the next upper layer. Together with the
coefficients of Eq. (6) (stored in the forward process), we obtain
the coefficients $a_l$ and $b_l$ of Eq. (7). The computation of
$a_l$ and $b_l$ and the solution of Eq. (7) together with SE
equation (2) to get a new source function $S(\tau _l)$ are
performed during the backward process layer by layer to the
surface. The process is iterated to the convergence.

\section{Applications of FBILI}

The accuracy and efficiency of the FBILI me\-thod have been
checked in several RT problems. Here, we list some of the
applications.

\subsection{Two-Level-Atom Line Formation}

The method was first developed for the two-level-atom line
formation (with complete redistribution and no overlapping
continuum) in a plane-parallel constant property medium (for
details see \citeauthor{av91} \citeyear{av91}, \citeauthor{acs97}
\citeyear{acs97}). This case represents an ideal test for checking
the numerical accuracy and stability of a new method. Namely,
under such conditions the features of the solution depend only on
the NLTE parameter $\varepsilon $ that is usually very small, so
that the numerical errors can easily blur the solution.

Thus, in order to test the stability of the method we solved the
two-level atom problem with $\varepsilon = 10^{-12}$. The results
are given in Fig. 1. They are compared with the exact
discrete-ordinate solution \citep{ah65}, obtained by using the
same discretization in optical depth (10 points per decade). The
asymptotic value of the maximum relative error of the order of
0.3$\%$ is reached already within 9-14 iterations. Namely, only
nine iterations are sufficient for the maximum relative correction
between two successive iterations (for all depth points) to be
less than $\delta =10^{-2}$ and 14 iterations for $\delta
=10^{-3}$. Hence, we see that a negligible additional effort (of
the iterative computation of the coefficients of the {\it
implicit} relations, instead of the mean intensities themselves)
with respect to the classical $\Lambda $ iteration results in an
extremely fast convergence (about 10 FBILI iterations compared to
about $1/\varepsilon $ classical $\Lambda $ iterations, while one
FBILI iteration takes only about 10$\%$ more CPU time than a
classical $\Lambda $ iteration).

In the literature the performances of various methods are usually
given for $\varepsilon =10^{-4}$. Thus, for this case, in Fig. 2
the convergence properties of the FBILI method are compared with
those of the ALI methods that use diagonal and 3-diagonal
approximate $\Lambda ^*$ operators \citep{hub03}. Excellent
convergence properties of the FBILI are evident. Using FBILI
yields convergence that is comparable to or even faster (for
$\delta \ge 10^{-3}$) than the 3-diagonal operator with Ng
acceleration!

\begin{figure}[!ht]


\plotfiddle{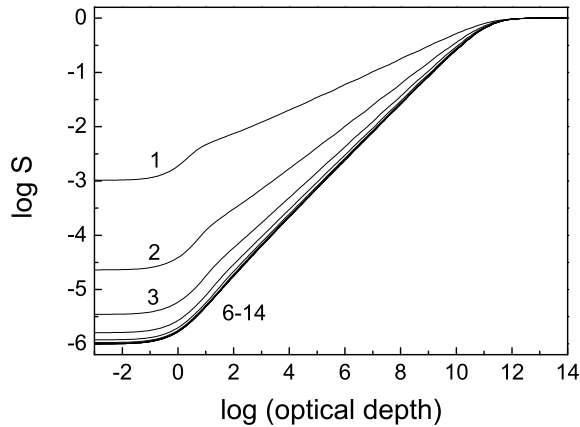}{5cm}{0}{80}{80}{-150}{-15}


\caption{Evolution with iterations of the source function for a
two-level atom with $\varepsilon = 10^{-12}$ and $B=1$.}

\end{figure}

\begin{figure}[!ht]


\plotfiddle{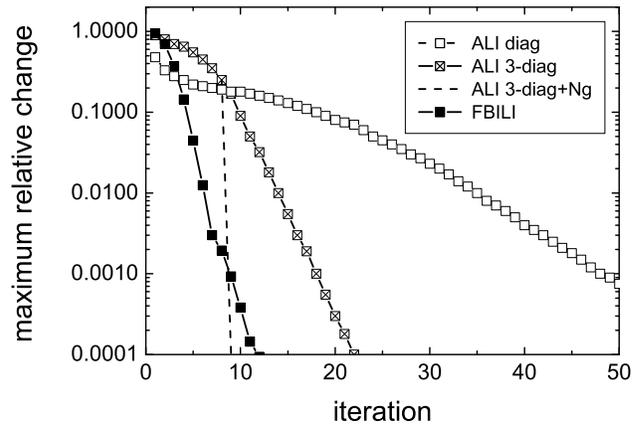}{5cm}{0}{80}{80}{-150}{-20}


\caption{Maximum relative change $\delta $ of the source function
as a function of the iteration number, for different variants of
the ALI scheme (reproduced from Fig. 1 in \citeauthor{hub03}
\citeyear{hub03}) and for the FBILI method. In all cases,
two-level atom without continuum is considered, with $\varepsilon
=10^{-4}$, $B=1$ and 4 depth-points per decade of optical depth.}

\end{figure}

\subsection{Two-Level-Atom Line Formation: Partial Redistribution}

The FBILI method was applied to the case of the two-level atom
line formation problem in which partial redistribution is taken
into account in the paper by \citet{acs97}. The results reproduced
the well-known ones by \citet{humm69}. For the cases $\varepsilon
=10^{-4}$ and $\varepsilon =10^{-8}$, 13 and 15 iterations,
respectively, are enough to fulfill the criterion $\delta
=10^{-3}$ for all the frequencies and all optical depths.

\subsection{Multilevel-Atom Line Formation}

The procedure for multilevel problem is the same as in the
two-level-atom case. Starting with the known set of level
populations, we repeat the entire forward process for each
radiative transition $i\rightarrow j$. In the backward process,
layer by layer, we compute the coefficients of the linear relation
(7) for all the transitions, and replacing them in the SE
equations we solve the latter for the new set of level
populations. The test is performed by solving the same problem
(three-level hydrogen atom line formation in an isothermal
atmosphere) as in \citet{al87}. The solution with a maximum
relative error below 3$\%$ is obtained in only nine iterations
with the convergence criterion $\delta =10^{-3}$
\citep[see][]{acs97}.

\subsection{Spherical Radiative Transfer: Monochromatic Scattering}

The generalization of the FBILI method to spherical geometry is
performed by \citet{av03}. Monochromatic scattering problem in a
spherical atmosphere is solved and the results are compared with
those given by \citet{gros} and \citet{al84}. The relative
difference of the solutions is about 1$\%$. The solution is
obtained already in the second iteration, whereas three iterations
are required for the maximum relative correction to be less than
1$\%$.

\subsection{Spherical Radiative Transfer: Line Formation}

In order to test the feasibility of the method when applied to the
line formation in spherically symmetric media, the test problem of
the line transfer with background absorption, proposed by
\citet*{mkh75} and \citet{al87}, is solved. The solution is
obtained in 15 iterations with an error less than 2$\%$.

\section{Conclusions}

Forth-and-Back Implicit Lambda Iteration method is a simple,
stable and extremely fast convergent iterative method developed
for the exact solution of NLTE RT problems. A negligible
additional computational effort with respect to the classical
$\Lambda $ iteration results in an extremely fast convergence. No
additional acceleration is needed. The method is easy to apply. No
matrix formalism is required so that the memory storage grows
linearly with dimension of the problem. Due to its great
simplicity and considerable savings in computational time and
memory storage FBILI seems to be a far-reaching tool to deal with
more complex problems (multidimensional RT, RT in moving media, or
when RT has to be coupled with other physical phenomena).

\acknowledgements I would like to thank the organizers of the
meeting and UNESCO-ROSTE for financial support. I would also like
to thank prof. P. Heinzel for inspiring discussions that helped to
improve the paper. This work has been realized within the Project
No.146003G supported by the Ministry of Science and Environmental
Protection of the Republic of Serbia.

\end{document}